\documentclass[proceedings, preprint]{rmaa}

\usepackage{paralist}

\usepackage{psfrag,color}

\SetYear{2015}
\SetConfTitle{Fourth Workshop on Robotic Autonomous Observatories}

\title{Astrophysics of ``extreme'' solar-like stars} 

\author{
  M. D. Caballero-Garc\'{i}a,\altaffilmark{1} 
  A. J. Castro-Tirado,\altaffilmark{2}
  A. Claret,\altaffilmark{2}
  K. Gazeas,\altaffilmark{3}
  V. \v{S}imon,\altaffilmark{4,5}
  M. Jel\'{\i}nek,\altaffilmark{4}
  A. Cwiek,\altaffilmark{6}
  A. F. \.Zarnecki,\altaffilmark{7}
  S. Oates,\altaffilmark{2}
  S. Jeong,\altaffilmark{2}
  and R. Hudec\altaffilmark{4,5}}

\altaffiltext{1}{Astronomical Institute, Academy of Sciences of the Czech 
Republic, Bo\v{c}n\'{\i}~II 1401, CZ-141\,00~Prague, Czech Republic (garcia@asu.cas.cz).}
\altaffiltext{2}{Instituto de Astrof\'{\i}sica de Andaluc\'{\i}a (IAA-CSIC), P.O. Box 03004, E-18080, Granada, Spain.}
\altaffiltext{3}{University of Athens, Department of Astrophysics, Astronomy and Mechanics, Zografos, Athens, GR 15784, Greece.}
\altaffiltext{4}{Astronomical Institute, Academy of Sciences of the Czech Republic, 251~65~Ond\v{r}ejov, Czech Republic.}
\altaffiltext{5}{Czech Technical University in Prague, Faculty of Electrical Engineering, Technick\'a 2, 166 27  Praha 6, Czech Republic.}
\altaffiltext{6}{National Centre for Nuclear Research, Ho\.za 69, 00-681 Warsaw, Poland.}
\altaffiltext{7}{Faculty of Physics, University of Warsaw, Pasteura 5, 02-093 Warszawa, Poland.}

\altaffiltext{3}{Please note that affiliations end in periods.}
\altaffiltext{4}{Full postal addresses are now given here instead of
  at the end of the paper (new as of v3.19, 19 Jan 2006).}

\shortauthor{Caballero-Garc\'{i}a et~al.}
\shorttitle{``Extreme'' solar-like stars}

\abstract{Only a few red dwarf flaring stars in the solar neighbourhood have undergone exceptional events called superflares. They have
been detected with high-energy satellites ({\it Swift}) and have been proven to be powerful events (both in intensity and energy) and potentially
hazardous for any extraterrestial life. The physics of these events can be understood as an extrapolation of the (much) weaker activity
already occurring in the most powerful solar flares occurring in the Sun. Nevertheless, the origin (why?) these superflares occur is
currently unknown. A recent study presents the optical and X-ray long-term evolution of the emission by the super-flare from the red-dwarf star DG~CVn undertaken in 2\,014. In that paper we comment on the context of these observations and on the properties that can be derived through the analysis of them.  }

\resumen{S\'olo unas pocas estrellas enanas rojas en la vecindad solar han experimentado eventos excepcionales llamados s\'uper-erupciones. \'Estas han sido 
detectadas por los sat\'elites de alta energ\'ia (entre ellos el sat\'elite {\it Swift} de la NASA) y se ha demostrado que son acontecimientos de gran alcance (tanto en 
intensidad como en energ\'ia) y potencialmente
peligrosos para cualquier tipo de vida extraterrestre. La f\'isica de estos eventos puede ser entendida como una extrapolaci\'on de la actividad (mucho) m\'as d\'ebil
de la que ocurre en las llamaradas m\'as potentes que se producen en el Sol. Sin embargo, el origen (el por qu\'e?) estas s\'uper-llamaradas ocurren 
actualmente se desconoce. Un estudio reciente presenta la emisi\'on \'optica y de rayos X mas la evoluci\'on a largo plazo de la s\'uper-llamarada de la estrella enana roja DG~CVn 
que tuvo lugar en 2\,014. En dicho art\'iculo se comenta acerca del contexto de estas observaciones y de las propiedades que se pueden derivar a trav\'es del an\'alisis de las mismas. 
}

\addkeyword{Gamma-rays: stars}
\addkeyword{Stars: flare}
\addkeyword{Stars: activity}

\begin{document}
\maketitle

\section{Introduction}
\label{KK}

Recently, the study of stars similar to the Sun (i.e. those in or close to the main sequence), including its younger and/or older counterparts, is 
a topic of ``top'' research. Also, thanks to the discovery of new extra-terrestial 
planets (mainly through the discoveries from the {\it KEPLER} mission and others) scientists are currently getting new insights into the importance of the interaction between
stars and their respective planets. This interaction may be so important that it might constitute one of the main drivers into the existence of life in those planets. Not so
far from the Sun, we know about the role of the Sun-Earth interaction. Solar storms and powerful ejections in the direction to the Earth considerably affect our daily 
communications. These powerful solar ejections are due to to its magnetic field lines (originated by the solar ``dynamo'') that sometimes reach its external surface and create
these powerful energetic events through the so called ``magnetic reconnection'' process. It is only thanks to the magnetosphere of the Earth that we are protected from this ejection of
particles and radiation (in X-rays and optical) from the Sun. At least it has been like this during the last hundredth of years (i.e. where the activity of the Sun has been monitored
so far). Nevertheless, in other (more active) stars than the Sun this natural barrier might not be enough. They might experience such large ejection events that would 
decrease any chance of possibility of life in their hosted extra-terrestial planets to zero. This is a scientific case worth to explore (both in the X-ray and the optical 
energy ranges), in order to give us insights into the importance of the 
stellar activity and its influence in their boundaries.

The main goal of the X-ray {\it Swift} satellite \citep{gehrels04}
is the detection of Gamma-Ray Bursts (GRBs) due to its high angular field of view and rapid (time) reaction. Nevertheless, it has also been revealed to be a fantastic tool for the study
of the X-ray emission from transient sources (with often an unpredictable emission behaviour). Observations in the optical are performed by big and medium-sized telescopes on Earth. The
former are not suitable for performing the rapid follow-up needed for the study of optical transients (as we will explain hereafter). These transients events are typically of
short duration (from fractions of a second to a few days), because the physical processes that originate them are of limited duration/spatial extent. Robotic smaller telescopes are
very well suited for performing such studies. This is due to several factors: their observing flexibility, their rapid response and
slew times and the fact that they can be located worldwide working remotely (therefore allowing continuous monitoring). Additional
observations are triggered after the transient has been detected with large X-ray/Optical Observatories. In this way we can perform
deep studies on the nature of these sources.

\subsection{The Burst Optical Observer and Transient Exploring System and its Spectrographs}

{\it BOOTES} (acronym of the Burst Observer and Optical Transient Exploring System) is a world-wide network of robotic telescopes. It
was originally designed from a Spanish-Czech collaboration that started in 1998 \citep{castro99,castro12}. The telescopes are located
at Huelva ({\it BOOTES}--1), M\'alaga ({\it BOOTES}--2), Granada,
Auckland ({\it BOOTES}--3), Yunnan ({\it BOOTES}--4) and San Pedro M\'artir ({\it BOOTES}--5), located at Spain, New Zealand, China and Mexico, respectively. There are plans of extending
this network even further (South Africa, Chile,...). These
telescopes are medium-sized (${\rm D}=30-60$\,cm), autonomous and very versatile. They are very well suited for the continuous study of the fast variability from
sources of astrophysical origin (GRBs and Optical Transients).

Currently two spectrographs are built and working properly in the network at M\'alaga and Granada (in the optical and infra-red, respectively). In
the following Sections we will show preliminary results obtained so far with {\it COLORES} at {\it BOOTES}--2.

\subsection{{\it COLORES}}

{\it COLORES} stands for Compact Low Resolution Spectrograph \citep{rabaza14}. It is a spectrograph designed to be light-weight enough to be carried by the high-speed robotic
telescope 60\,cm ({\it BOOTES}--2). It works in the wavelength range of ($3\,800-11\,500$)\,{\AA} and has a spectral resolution of ($15-60$)\,{\AA}. The
primary scientific target of the spectrograph is a prompt GRB follow-up, particularly the estimation of redshift.

{\it COLORES} is a multi-mode instrument that can switch from imaging a field (target selection and precise pointing) to spectroscopy by rotating wheel-mounted
grisms, slits and filters. The filters and the grisms (only one is mounted at the moment) are located in standard filter wheels and the optical design is
comprised of a four-element refractive collimator and an identical four-element refractive camera. As a spectroscope, the instrument can use different
slits to match the atmospheric seeing, and different grisms in order to select the spectral resolution according to the need of the observation.

The current detector is a $1\,024{\times}1\,024$ pixels device, with 13 micron pixels. The telescope is a rapid and light-weight design, and a low instrument weight was
a significant constraint in the design as well as the need to be automatic and autonomous. For further details on description, operation and working with
{\it COLORES} we refer to M. Jelinek PhD thesis (and references therein).

\section{The red-dwarf star DG~CVn}
\label{KK}

\noindent DG Canum Venaticorum (DG~CVn; with coordinates (J2000) ${\alpha}=13^{\rm h}31^{\rm m}46.7^{\rm s}$, ${\delta}=29^{\circ}16^{'}36^{''}$; also
named G~165-8AB; \citealt{gliese91} and 1RXS~J133146.9$+$291631; \citealt{zickgraf03}) is a bright (${\rm V}=12.19$; \citealt{xu14}) and
close (${\rm D}=18$\,pc) visual dM4e binary system \citep{riedel14,henry94,delfosse98}. It is also a radio emitting source \citep{helfand99}.
It has been seen that the system has Ca, H and K lines in emission \citep{beers94}. The components
have an angular separation in the sky of $0.17$\,arcsec \citep{beuzit04}, an orbital period of ${\rm P}{\approx}7$\,yr and magnitudes ${\rm V}=12.64,12.93$, for the primary and
the secondary components, respectively \citep{riedel14,weis91}. It is classified as a (joint spectral type from unresolved multiples) M4.0V spectral-type star \citep{riedel14}. It is
also a red high proper motion dwarf-star (${\pi}{\approx}80\,{\rm mas}$; \citealt{jimenez12}).

One of the components of DG~CVn (it is not known which one) has been reported to be chromospherically active \citep{henry94} and one of the only three known M dwarf
ultra-fast rotators in the solar neighborhood. The projected rotational velocity
is $v{\sin(i)}=55.5\,{\rm km}\,{\rm s}^{-1}$, as measured from rotational broadening of the H emission lines seen in high-resolution
spectra \citep{delfosse98}. Recently it has been reported the detection of intense radio emission (the highest ever detected in an active red-dwarf) coinciding
with the time of the first and the second flares reported
in this paper \citep{fender15}. Because the two components of the system are separated by 3.6\,AU (${\approx}2\,500\,{\rm R}_{\star}$) the two stars are not magnetically interacting. Therefore
the intense radio emission has been interpreted as a consequence of the processes occurring in one of the stars (we will be referring to this as DG~CVn hereafter). For this
star to rotate so rapidly
a tertiary close companion (i.e. apart from the distant known companion) is expected to exist, but recent studies \citep{fender15} indicate that this might not be the case and that the
youth of the system is the cause. Therefore, this system is considered to be a young star ($30\,{\rm Myr}$; \citealt{riedel14,delfosse98}). Nevertheless, we refer to \citet{cabal15} for a discussion on its age.

\section{The optical and X-ray activity during the super-flare in 2014}
\label{KK}

On April 23rd 2014, at 21:07:08 UT one of the stars from DG~CVn flared bright enough (300
milliCrab in the 15-150\,keV band) to trigger the {\it Swift}
satellite's \citep{gehrels04} Burst Alert Telescope (BAT; \citealt{barthelmy05}).
Within two minutes of this ${T}_{0}$, {\it Swift} had slewed to point its
narrow-field telescopes to the source, which revealed gradually decreasing soft
X-ray emissions with a second weaker flare occurring at ${\rm T}_{0}+11$\,ks,
followed by several smaller flares. This behaviour was observed in both the optical
and X-ray bands. On the ground, the wide-field ``Pi of the Sky'' \citep{cwiok07,mankiewicz14} (PI) instrument was observing, covering {\it
Swift}'s field of view, and recorded the optical behaviour of DG~CVn even
before the burst began and continued until ${\rm T}_{0}+1100$\,s.
The {\it BOOTES}--2 \citep{castro99,castro12} telescope began to observe DG~CVn from ${\approx}{\rm
T}_{0}+11$\,min, starting to take spectra later at ${\approx}{\rm T}_{0}+1$\,h with the low-resolution spectrograph COLORES \citep{rabaza14}, covering the period of
the second flare. Observations with {\it BOOTES}--2 continued for several weeks following
the trigger. Deeper spectra were obtained later with instruments/spectrographs on larger telescopes: OSIRIS \citep{cepa00}
on the {\it Gran Telescopio de Canarias} (GTC) at ${\rm T}_{0}+1.2$\,d,
SCORPIO \citep{afanasiev05} on the 6\,m BTA-6 telescope at SAO in the Caucasus at ${\rm
T}_{0}+15$\,d and CAFE \citep{aceituno13} on the 2.2\,m telescope at Calar Alto at ${\rm
T}_{0}+53$\,d. Fig.~1 shows the data from the first flare: the optical lightcurve from PI
together with the 15-25\,keV {\it Swift} data, from ${\rm T}_{0}-150$\,s to
${\rm T}_{0}+300$\,s. Optical and X-ray observations by {\it Swift} XRT of the second flare are shown in Fig.~2.

\subsection{Observations with ``Pi of the Sky''}

\noindent The ``Pi of the Sky'' (PI) experiment is designed to monitor a large fraction of the sky with a high time resolution (10\,s) and self-triggering
capabilities \citep{cwiok07,mankiewicz14}. This means that PI may be performing observations of the field of sources well before the trigger time (${\rm T}_{0}$), the latter given by high-energy instruments (like {\it Swift}/BAT). This
approach resulted in the optical monitoring of DG~CVn even before ${\rm T}_{0}$. PI observed DG~CVn from ${\rm T}_{0}-700$\,s until ${\rm T}_{0}+1\,100$\,s. In
Fig.~1 the light curve is shown since ${\approx}{\rm T}_{0}-150$\,s onwards (since no significant brightness variations were detected before).

As can be seen, the first optical flash only lasted about 60\,s, with the source brightening by over 4\,mag, to ${\rm V}{\approx}7$ at maximum
(occurring at ${\rm t}={\rm T}_{0}-41.3{\pm}0.4$\,s), and ending before the maximum of the hard X-ray emission which triggered the BAT alert (${\rm T}_{0}$). Then, a slow
decrease of the optical brightness was observed reaching ${\rm V}=11$ after about 0.5\,h.


\section{Results}
\label{KK}

\citet{cabal15} studied the spectral evolution of DG~CVn during an episode of important optical and hard X-ray activity from 23rd April 2014 onwards. During this period at least
two flares have been detected
to occur quasi-simultaneously in both energy bands. It was explained in that paper that the origin of the (few tens of seconds) hard X-ray delay with respect to the optical emission (Fig. 1) by the Neupert
effect \citep{neupert68}, which was observed before (apart from the Sun) in UV Ceti and Proxima Centauri during normal
soft X-ray flares. The novelty of that study is that this effect has been observed, but for the hard X-ray emission.

The X-ray luminosities during the peaks of emission are
${\rm L}_{\rm X}(0.3-10)\,{\rm keV}=1.4{\times}10^{33},3.1{\times}10^{32}\,{\rm erg}\,{\rm s}^{-1}$. Although the flare X-ray luminosities are certainly high
similar values (${\rm L}_{\rm X}(0.3-10)\,{\rm keV}{\ge}10^{32}\,{\rm erg}\,{\rm s}^{-1}$) have been detected for a dozen of cases
in main-sequence red-dwarf stars. Recently these values have been superseded by an intense episode of activity from the
red-dwarf binary system SZ~Psc \citep{drake15}. Also the duration of the flaring episode is no exception, with similar values reported for the
(9\,day) flare of CF~Tuc observed by \citet{kurster96} and the previous super-flare from EV~Lac \citep{osten10}.

\begin{figure}[!t]
  \includegraphics[angle=270,bb=0 0 612 792,clip,width=\columnwidth]{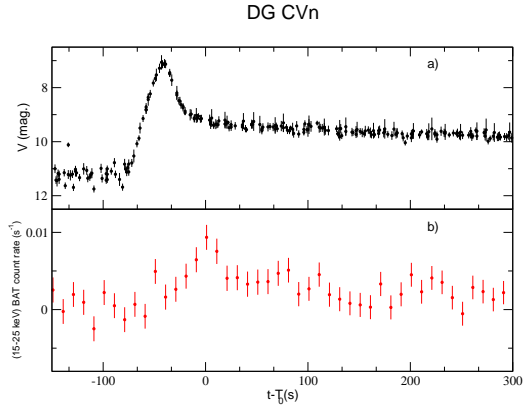}
  \caption{Hard X-ray delay versus the optical emission of the prompt emission (first flare) from DG~CVn. {\bf a}, Optical light
curve from the PI (from ${\rm T}_{0}-150$\,s onwards); {\bf b}, {\it Swift}/BAT light curve in the 15--25\,keV energy range (i.e. hard X-rays). The time (in seconds)
is measured with respect to the BAT trigger time (${\rm T}_{0}$; \citealt{delia14}). Figure taken from \citet{cabal15}.}
  \label{fig:crop}
\end{figure}

\begin{figure}[!t]
  \includegraphics[angle=270,bb=0 0 612 792,clip,width=\columnwidth]{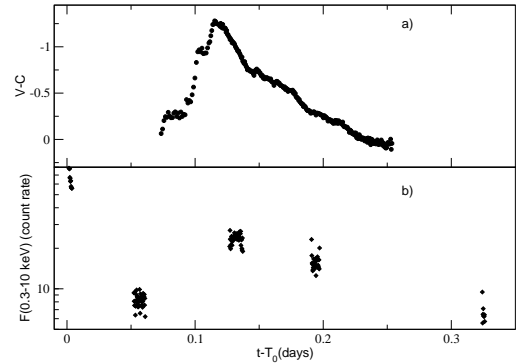}
  \caption{{\bf a}, Optical light curve variations (with respect to a photometric standard star) in V filter during the second flare
of DG~CVn at ${\rm t}-{\rm T}_{0}=0.127$\,d (credits: K. Gazeas); {\bf b}, {\it Swift}/XRT light curve in the 0.3--10\,keV energy range (the Flux is measured in ${\rm cts}\,{\rm s}^{-1}$). }
  \label{fig:crop}
\end{figure}

\section{Discussion and conclusions}
\label{KK}

Such episodes of increased X-ray/optical activity
have never been observed in the Sun so far. However, the same physics underlying the giant X-ray/optical solar flares
serves to explain the superflares observed in active-like stars (as DG~CVn; \citealt{aulanier13}). Nevertheless, since the most powerful solar flares have been of
only about $10^{32}\,{\rm erg}\,{\rm s}^{-1}$ \citep{carrington859} the scale-up of solar flares models would require enormous starspots (up to 48 degrees in latitude/longitude extent)
to match stellar superflares, thus much bigger than any sunspots in the last 4 centuries of solar observations. Therefore \citet{aulanier13} conjecture that one condition for
Sun-like stars to produce superflares is to host a dynamo that is much stronger than that of the Sun.

It is also worth to mention that some recent studies
\citep{kitze14,candelaresi14,wu15} point out to both the (possible) few instances and the possibility of solar-type stars undergoing superflares with luminosities as high as
$10^{35}-10^{37}\,{\rm erg}\,{\rm s}^{-1}$ under certain conditions, by using data from the {\it KEPLER} satellite \citep{koch10}. Therefore, this indicates the potential
of such events as powerful releases of energy. If so, that would decrease to zero the possibilities of existence of life in any possible hosted planet in these stellar systems.

\acknowledgments{We thank the scientific web portal http://cienciaes.com/ for the public outreach of this work. This 
publication was supported by the European social fund within the 
framework of realizing the project "Support of inter-sectoral mobility and quality 
enhancement of research teams at Czech Technical University in Prague“, CZ.1.07/2.3.00/30.0034.  }


\begin{thebibliography}
\bibitem[Aceituno et~al.(2013)]{aceituno13}
Aceituno, J., et~al., 2013, A\&A, 552, 31
\bibitem[Afanasiev et~al.(2005)]{afanasiev05}
Afanasiev, V.~L., Moiseev, A.~V., 2005, Astronomy Letters, 31, 194-204
\bibitem[Aulanier et~al.(2013)]{aulanier13}
Aulanier, G., D{\'e}moulin, P., Schrijver, C.~J., et~al., 2013, A\&A, 549, 66
\bibitem[Barthelmy et~al.(2005)]{barthelmy05}
Barthelmy, S.~D., Barbier, L.~M., Cummings, J.~R., et~al., 2005, SSRv, 120, 143
\bibitem[Beers, Bestman \& Wilhelm(1994)]{beers94}
Beers, T.~C., Bestman, W. \& Wilhelm, R., 1994, AJ, 108, 268
\bibitem[Beuzit et~al.(2004)]{beuzit04}
Beuzit, J.-L., S{\'e}gransan, D., Forveille, T., Udry, S., et~al., 2004, A\&A, 425, 997
\bibitem[Caballero-Garc{\'{\i}}a et~al.(2015)]{cabal15}
Caballero-Garc{\'{\i}}a, M.~D., {\v S}imon, V., Jel{\'{\i}}nek, M., Castro-Tirado, A.~J., Cwiek, A., Claret, A., et~al., 2015, MNRAS, 452, 4195
\bibitem[Candelaresi et~al.(2014)]{candelaresi14}
Candelaresi, S., Hillier, A., Maehara, H., et~al., 2014, ApJ, 792, 67
\bibitem[Carrington(1859)]{carrington859}
Carrington, R.~C., 1859, MNRAS, 20, 13
\bibitem[Castro-Tirado et~al.(1999)]{castro99}
Castro-Tirado, A.~J., Sold{\'a}n, J., Bernas, M., et~al., 1999, A\&AS, 138, 583
\bibitem[Castro-Tirado et~al.(2012)]{castro12}
Castro-Tirado, A.~J., Jel{\'{\i}}nek, M., Gorosabel, J., et~al., 2012, ASI Conf. Ser. 7, 313
\bibitem[Cepa et~al.(2000)]{cepa00}
Cepa, J., et~al., Optical and IR Telescope Instrumentation and Detectors, Eds. Iye, M. \& Moorwood, A.~F., 2000, 4008, 623-631
\bibitem[{\'C}wiok et~al.(2007)]{cwiok07}
{\'C}wiok, M., Dominik, W., Ma{\l}ek, K., Mankiewicz, L. et~al., 2007, Ap\&SS, 309, 531
\bibitem[D'Elia et~al.(2014)]{delia14}
D'Elia, V., Gehrels, N., Holland, S.~T., Krimm, H.~A., et~al., 2014, GCN, 16158, 1
\bibitem[Delfosse et~al.(1998)]{delfosse98}
Delfosse, X., Forveille, T., Perrier, C. \& Mayor, M., 1998, A\&A, 331, 581
\bibitem[Drake et~al.(2015)]{drake15}
Drake, S., Osten, R., Krimm, H., et~al., 2015, ATel, 6940, 1
\bibitem[Fender et~al.(2015)]{fender15}
Fender, R.~P., Anderson, G.~E., Osten, R., et~al., 2015, MNRAS, 446L,66
\bibitem[Gehrels et~al.(2004)]{gehrels04}
Gehrels, N., Chincarini, G., Giommi, P., et~al., 2004, \apj, 611, 1005
\bibitem[Gliese \& Jahrei{\ss}(1991)]{gliese91}
Gliese, W. \& Jahrei{\ss}, H., 1991, Selected Astronomical Catalogs, Vol. I; L.E. Brotzmann, S.E. Gesser (eds.)
\bibitem[Helfand et~al.(1999)]{helfand99}
Helfand, D.~J., Schnee, S., Becker, R.~H., White, R.~L. \& McMahon, R.~G., 1999, AJ, 117, 1568
\bibitem[Henry, Kirkpatrick \& Simons(1994)]{henry94}
Henry, T.~J., Kirkpatrick, J.~D., Simons, D.~A., 1994, AJ, 108, 1437
\bibitem[Jim{\'e}nez-Esteban et~al.(2012)]{jimenez12}
Jim{\'e}nez-Esteban, F.~M., Caballero, J.~A., Dorda, R., Miles-P{\'a}ez, P.~A., et~al., 2012, A\&A, 539A, 86
\bibitem[Kitze et~al.(2014)]{kitze14}
Kitze, M., Neuh{\"a}user, R., Hambaryan, V. \& Ginski, C., 2014, ApJS, 77, 417
\bibitem[Koch et~al.(2010)]{koch10}
Koch, D.~G., Borucki, W.~J., Basri, G., et~al., 2010, ApJL, 713, 79
\bibitem[Kurster \& Schmitt(1996)]{kurster96}
Kurster \& Schmitt, 1996, A\&A, 311, 211 
\bibitem[Mankiewicz et~al.(2014)]{mankiewicz14}
Mankiewicz, L., et~al., RevMexAA (SC), 45, 2014, 7-11
\bibitem[Neupert(1968)]{neupert68}
Neupert, W.~M., 1968, ApJ, 153L, 59
\bibitem[Osten et~al.(2010)]{osten10}
Osten, R.~A., Godet, O., Drake, S., Tueller, J., et~al., 2010, ApJ, 721, 785
\bibitem[Rabaza et~al.(2014)]{rabaza14}
Rabaza, O., Jelinek M., Castro-Tirado A.~J., et~al., 2014, Review of Scientific Instruments 84 (11), 114501
\bibitem[Riedel et~al.(2014)]{riedel14}
Riedel, A.~R., Finch, C.~T., Henry, T.~J., Subasavage, J.~P., et~al., 2014, AJ, 147, 85
\bibitem[Weis(1991)]{weis91}
Weis, E.~W., 1991, AJ, 102, 1795
\bibitem[Wu et~al.(2015)]{wu15}
Wu, C.-J., Ip, W.-H. \& Huang, L.-C., 2015, ApJ, 2015, 798, 92
\bibitem[Xu et~al.(2014)]{xu14}
Xu, D., Bai, C.-H., Zhang, X. \& Esamdin, A., 2014, GCN, 16159, 1
\bibitem[Zickgraf et~al.(2003)]{zickgraf03}
Zickgraf, F.-J., Engels, D., Hagen, H.-J., Reimers, D. \& Voges, W., 2003, A\&A, 406, 535





































\end{thebibliography}
\end{document}